\documentclass[12pt,a4paper]{article}
\usepackage{cite,color,graphics,amsmath,epsfig,rotating}
\usepackage{siunitx}
\pdfoutput=1
   
\usepackage{latexsym}
\usepackage{amssymb}
\usepackage{slashed}
\usepackage{mathrsfs}
\usepackage{graphicx}
\usepackage{subfig}
\usepackage{psfrag}
\usepackage{url}
\DeclareMathAlphabet{\mathpzc}{OT1}{pzc}{m}{it}

\usepackage{hyperref}
\hypersetup{colorlinks=true,linktocpage,bookmarksopen,bookmarksnumbered,pdfstartview=FitH,breaklinks=true}
\ifx\pdfoutput\undefined

\usepackage[bookmarks]{hyperref}        
\usepackage{enumitem}
\else
\usepackage{hyperref}   
\fi

\usepackage{multicol}
\usepackage{color}
\usepackage{marginnote}
\usepackage{eurosym}

\def\micro{{\tt micrOMEGAs}}

\begin{document}
\begin{center}

{\bf WIMP and FIMP dark matter in the inert doublet plus singlet model.}

\vspace*{1cm}\renewcommand{\thefootnote}{\fnsymbol{footnote}}

{\large  G.~B\'elanger$^{1}$, 
A.~Mjallal$^{1}$, 
A.~Pukhov$^{2}$

\renewcommand{\thefootnote}{\arabic{footnote}}

\vspace*{1cm} 
{\normalsize \it 
$^1\,$ \href{http://lapth.cnrs.fr}{LAPTh}, CNRS, USMB, 9 Chemin de Bellevue, 74940 Annecy, France\\[2mm]
$^2\,$\href{http://theory.sinp.msu.ru}{Skobeltsyn Institute of Nuclear Physics}, Moscow State University,\\ Moscow 119992, Russia\\[2mm]
}}

\vspace{1cm}

\begin{abstract}
We consider multi-component dark matter in a model where one dark matter component is feebly interacting (FIMP)  while the second is weakly interacting (WIMP). 
The model contains an inert scalar doublet and a complex scalar singlet and features a discrete $Z_4$ symmetry. 
We determine the parameter space that satisfies  theoretical constraints,   collider constraints, relic density as well as direct and indirect detection limits. We discuss the possibility to probe the model through  collider and astrophysical searches. We find that it is possible that the FIMP forms the dominant DM component, in this case the astrophysical signatures of the WIMP are much suppressed. 
We also explore the case where the doublet can decay into pairs of singlets leaving only one DM component, the FIMP. These scenarios are constrained by BBN and are best explored at colliders since astrophysical signatures are suppressed. 
  \end{abstract} 

\end{center}

\section{ Introduction}

While there is strong evidence that a large fraction of the matter content of the Universe is dark, the nature of this dark matter (DM)  remains a mystery. Potential candidates for a new stable and neutral particle are predicted in extensions of the standard model. The most studied case is the one of a new weakly interacting particle which naturally reaches the measured value of the relic density through the freeze-out mechanism~\cite{Arcadi:2017kky,Roszkowski:2017nbc}. Search strategies that have been designed for WIMPs in astroparticle physics and at colliders have not yet found evidence of a new DM particle~\cite{Aprile:2018dbl,Fermi-LAT:2015att,Fermi-LAT:2016uux,HAWC:2019jvm,Abercrombie:2015wmb,PandaX-II:2020oim,PICO:2019vsc,Fermi-LAT:2019lyf}. Many other possibilities have been entertained both from the point of view of particle model-building and  dark matter production mechanisms in the early Universe. Such mechanisms  include  freeze-in production of feebly interacting particles~\cite{McDonald:1993ex,Hall:2009bx,Bernal:2017kxu},  asymmetric dark matter~\cite{Petraki:2013wwa,Zurek:2013wia}, as well as strongly interacting dark matter~\cite{Hochberg:2014dra,Hochberg:2014kqa,Lee:2015gsa,Smirnov:2020zwf}. In these scenarios the search strategies can differ significantly from the WIMP case. For example FIMPs can be significantly lighter than the typical WIMP and typically feature negligible cross-section for elastic scattering on nuclei, thus escaping strong direct detection constraints. FIMPs can nevertheless be tested in direct detection when they are coupled to a very light  mediator~\cite{Hambye:2018dpi,Belanger:2020npe}.  Moreover due to the feeble couplings, most of the collider searches for missing energy signature have very low rates for  FIMPs. In some cases though new collider signatures can emerge, for example when the FIMP is accompanied by a charged particle in the dark sector, signatures such as displaced vertices can become relevant~\cite{Hessler:2016kwm,Belanger:2018sti,Calibbi:2018fqf,Brooijmans:2020yij,Calibbi:2021fld}. 

Although a single new stable particle can easily explain dark matter observables, there is {\it a priori} no reason to believe that the dark sector contains only one particle nor that only one of these is stable. Multi-component dark matter models thus provide a viable extension of the simplest dark matter models that allow to escape some of the strong constraints put on WIMPs from the lack of confirmed signals in direct detection, indirect detection and colliders. This occurs mainly because it is possible in such models to uncorrelate  the processes responsible for direct detection (or indirect detection) with the ones entering dark matter formation.
Such a situation can occur in models with two WIMPs, when  the component that has a large elastic cross-section on nuclei forms only a subdominant DM component leading to a suppressed  signal.
Moreover  
interactions between the two dark sectors can also suppress the relic density thus requiring weaker interactions of DM with the SM with the consequence of relaxing various constraints ~\cite{Belanger:2014bga}. Escaping strong experimental constraints seems even more natural when the dominant DM component is a FIMP while the WIMP is subdominant thus leading to suppressed signal in both direct and indirect searches. However signals at colliders are independent of the fraction of each DM component in the universe and those for the WIMP are therefore not suppressed in these scenarios.

In a previous publication~\cite{Belanger:2021lwd} we have examined carefully a model with two dark sectors,
the inert doublet and singlet model (IDSM) which contains both  a doublet and a singlet of scalars in addition to the standard model (SM) particles. The model also  features  a discrete $Z_4$ symmetry where the singlet and doublet have different charges~\cite{Belanger:2012vp,Belanger:2014bga}. This  guarantees the stability of the singlet and of the lightest neutral component of the doublet if its decay to two singlets is kinematically forbidden. Within this framework we have shown how current constraints that affect both the inert doublet and singlet models that form  subsets of the IDSM  can be significantly relaxed notably because of the interactions between the two sectors~\cite{Belanger:2021lwd}. In this publication we consider the same model but assume that the singlet is feebly interacting with the standard model such that it is not in thermal equilibrium in the early universe. We also impose the condition that the singlet is feebly interacting with the inert doublet, otherwise  it would be brought to  thermal equilibrium with the standard model  through the inert doublet. We explore how  the model can reproduce the total dark matter relic density  while evading constraints on WIMP dark matter  and discuss possible probes of the model. One important distinctive feature of having a FIMP is that its mass can be significantly below the weak scale.
We therefore examine carefully cases where the decay of the doublet into singlets is kinematically allowed, leading to only one DM. In such scenarios we find that both Big Bang Nucleosynthesis (BBN) and the Cosmologigal Microwave Background (CMB) constrain the cases with a long-lived doublet.   
 
The paper is organised as follows. The model is briefly reviewed  in Section 2  and the calculation of the relic density  is described in section 3. 
Section 4 reviews various constraints on the model. The results of scans over the parameter space of the model with two DM components are presented in section 5 and those for a single DM component in  section 6. Section 7 contains our conclusions.

\section{The model}
\label{model}

The model is an extension of  the SM  featuring a scalar sector which contains  in addition to  the SM scalar doublet Higgs $H$, an inert doublet $H'$ and a complex singlet $S$. All standard model particles are  invariant under a discrete  $Z_4$ symmetry while $H'$ and  $S$ transform non-trivially as  $\phi \to \exp^{(i X \pi/2)} \phi$
with the  charges $X_{H'}=2$ and $X_S=1$. The model  has two dark sectors, the first contains only the complex singlet while the second contains the doublet. 
The model  allows potentially for two dark matter candidates corresponding to  the lightest component of each sector.  
In general the scalar potential reads 
The scalar potential reads 
\begin{eqnarray}
V_{Z_4}&=&\lambda_1\left(\left|H \right|^2-\frac{v^2}{2}  \right)^2 + \mu_2^2\left|\tilde{H}'^2 \right| +\lambda_2\left| \tilde{H}'\right|^4 + \mu_S^2 \left|\tilde{S} \right| ^2 + \lambda_S \left| \tilde{S}\right|^4 +\frac{\lambda^\prime_S}{2}\left(\tilde{S}^4+\tilde{S}^{\dagger 4} \right) \nonumber\\
&&+ \lambda_{S1} \left|\tilde{S} \right|^2 \left| H\right| ^2 + \lambda_{S2} \left|\tilde{S} \right| ^2 \left|\tilde{H}' \right| ^2 \nonumber\\
&&+ \lambda_3 \left|H \right| ^2 \left| \tilde{H}'\right| ^2 + \lambda_4 \left(H^\dagger \tilde{H}' \right)\left(\tilde{H}'^\dagger H \right)     
+ \frac{\lambda_5}{2}\left[\left(H^\dagger \tilde{H}' \right)^2 + \left(\tilde{H}'^\dagger H \right)^2\right]  \nonumber\\&&+\frac{\lambda_{S12}}{2} \left( \tilde{S}^2H^\dagger \tilde{H}' + \tilde{S}^{\dagger 2} \tilde{H}'^\dagger H\right) + \frac{\lambda_{S21}}{2}\left(\tilde{S}^2 \tilde{H}'^\dagger H + S^{\dagger 2} H^\dagger \tilde{H}' \right)
\label{eq:VZ4}
\end{eqnarray}
where the components of the dark doublet are
\begin{equation}
\tilde{H}'=\begin{pmatrix}  -i \tilde{H}^+\\ \frac{\tilde{H}+ i \tilde{A}}{\sqrt{2}}  \end{pmatrix}
\end{equation}
Only the standard model doublet gets a vacuum expectation value, $v$.  We choose the five masses of the scalar fields as independent parameters,  $M_h=125 {\rm GeV}$  for the SM Higgs, $M_H,M_A,M_{H^\pm}$ for the two neutral and charged doublet and $M_S$ for the singlet. For the doublet, the mass parameters can also be replaced with the mass difference with DM,\footnote{We choose $\lambda_5>0$, hence the pseudoscalar component of the doublet is the DM. The DM phenomenology is expected to be similar if  the scalar component of the doublet is the DM. 
}
\begin{equation}
\Delta^+=M_{H^+} -M_A \;,\;\;\;\;  \Delta^0=M_{H} -M_A
\end{equation}
Five parameters of the potential can then  be expressed in terms of these masses,
\begin{eqnarray}
\lambda_1=\frac{M_h^2}{2v^2} \;;\;\;\;   \mu_2^2= M^2_{H^\pm}-\lambda_{3}\frac{v^2}{2}\;;\;\;\; \mu_S^2=M_S^2-\lambda_{S1}\frac{v^2}{2} \nonumber\\
\lambda_4 = \frac{M_H^2+M_A^2-2M_{H^\pm}^2}{v^2} \;;\;\;\; \lambda_5 = \frac{M_H^2-M_A^2}{v^2}
\end{eqnarray}
Amongst the remaining couplings that enter the potential,   $ \lambda_2,\lambda_S,\lambda'_S$ play little role in dark matter observables, For simplicity we will fix 
$ \lambda_S=1,\lambda'_S=1/2$ while we keep $\lambda_2$ as a free parameter because it plays a role in deriving theoretical constrains on the potential. The free parameters that will be considered then include 

\begin{equation}
M_H,M_A,M_{H^\pm}, M_S,  \lambda_2, \lambda_3,\lambda_{S1},\lambda_{S2},\lambda_{S12},\lambda_{S21}
 \label{eq:free}
\end{equation}

In the following we will consider only the case where the singlet is a FIMP, this means that the couplings $\lambda_{S1}, \lambda_{S2},\lambda_{S12}, \lambda_{S21}$ must all be feeble to prevent $\tilde{S}$ to be in thermal equilibrium with the standard model.  We will analyse both the case where only the FIMP is DM and the case where  two  stable particles form DM. The latter requires that   the decay of the neutral inert doublet into the singlet be kinematically forbidden,  that is $M_{A}< 2M_S$. 
 
The trilinear and quartic  scalar vertices that will play a role in DM observables are listed below:
\begin{center}
\begin{tabular}{llll}
$h\tilde{A} \tilde{A}$ &$-\frac{2M_W}{g} \lambda_{Ah}$ &\;\;\;$hh\tilde{A} \tilde{A}$ &$- \lambda_{Ah}$\nonumber\\
$h\tilde{H} \tilde{H} $&$-\frac{2M_W}{g} \lambda_{Hh}$&\;\;\;$hh\tilde{H} \tilde{H}$ &$- \lambda_{Hh}$\nonumber\\ 
$h\tilde{H}^+\tilde{H}^-$ & $-\frac{2M_W}{g} \lambda_3$&\;\;\;$hh\tilde{H}^+ \tilde{H}^-$ &$-\lambda_3$\nonumber\\
$h \tilde{S}\tilde{S}^\dagger$ &$-\frac{2M_W}{g} \lambda_{S1}$ &\;\;\; $hh\tilde{S} \tilde{S}^\dagger$ & $- \lambda_{S1}$    \nonumber\\
$\tilde{A} \tilde{S}\tilde{S}$ &$-i \frac{M_W}{g} (\lambda_{S21} -  \lambda_{S12})$ &\;\;\;$\tilde{A} \tilde{A}\tilde{S} \tilde{S}^\dagger$ &$- \lambda_{S2}$ \nonumber\\
&&\;\;\;$h\tilde{A} \tilde{S}\tilde{S}$ &$\frac{i}{2} (\lambda_{S21} -  \lambda_{S12})$\nonumber\\
$\tilde{H} \tilde{S}\tilde{S}$ &$-\frac{M_W}{g} (\lambda_{S21} +  \lambda_{S12})$&\;\;\;$\tilde{H} \tilde{H}\tilde{S} \tilde{S}^\dagger$ &$- \lambda_{S2}$ \nonumber\\
&&\;\;\;$h\tilde{H} \tilde{S}\tilde{S}$ &$-\frac{1}{2} (\lambda_{S21} +  \lambda_{S12})$\nonumber\\
&&\;\;\;$\tilde{H}^+ \tilde{H}^-\tilde{S} \tilde{S}^\dagger$ &$- \lambda_{S2}$ \nonumber\\
     \end{tabular}
     \end{center}
Here  $\lambda_{Hh}=\lambda_3+\lambda_4+\lambda_5$ and $\lambda_{Ah}=\lambda_3+\lambda_4-\lambda_5$ .

We have also included in the model  effective vertices for $\tilde{H}^\pm\tilde{A}\pi^\mp$ and $\tilde{H}^\pm\tilde{H}\pi^\mp$ interactions. These  vertices  become relevant when the mass splitting between the charged and neutral Higgses is below a few hundred MeV's~\cite{Chen:1996ap}. In this case the dominant decay mode is $\tilde{H}^\pm\rightarrow \pi^\pm \tilde{A}$ and  the decay width is much larger  than the one computed for decay into quarks $\tilde{H}^+\rightarrow  u\bar{d} A$. More details of the implementation can be found in ~\cite{Belyaev:2016lok,Belyaev:2020wok,Belanger:2021lwd}.

\section{Dark matter observables}
 \label{sec:relic}
 
 We first discuss dark matter observables and in particular the relic density of the two DM in the case of one FIMP (the singlet) and one WIMP (the doublet). 

\subsection{Relic density}
\label{sec:relic}

 In the following we split the particle content in different sectors according to their $Z_4$ charge.  The dark sector 1 contains the singlet, the dark sector 2 contains the doublet while all SM particles are in sector  $0$. The equations for the abundances of the FIMP (the first DM)  and the WIMP (the second DM) are written in the form 
 
  \begin{eqnarray}
3H\frac{dY_1}{ds}&=& \langle v\sigma^{1100}\rangle  \left(Y_1^2-\bar{Y}_1^2 \right) 
                  +     \langle v\sigma^{1120}\rangle\left( Y_1^2 - Y_2 \frac{\bar{Y}_1^2}{\bar{Y}_2} \right)
                                         + \langle v\sigma^{1122}\rangle\left( Y_1^2- Y_2^2  \frac{\bar{Y}_1^2}{\bar{Y}_2^2}\right)  \nonumber\\
                                         &-&2 \Gamma_{eff}(T)(Y_2 -Y_1^2\frac{\bar{Y}_2}{\bar{Y}_1^2})\nonumber\\
3H\frac{dY_2}{ds}&=&  \langle v\sigma^{2200}\rangle  \left(Y_2^2-\bar{Y}_2^2 \right) 
                   +  \langle v\sigma^{2211}\rangle\left(Y_2^2- Y_1^2 \frac{\bar{Y}_2^2} { \bar{Y}_1^2} \right)             
- \frac{1}{2} \langle v\sigma^{1120}\rangle\left(Y_1^2-Y_2\frac{\bar{Y}_1^2}{\bar{Y}_2}\right) \nonumber\\
&+& \langle v\sigma^{2110}\rangle\left( Y_1 Y_2- Y_1 \bar{Y}_2 \right)
  +\Gamma_{eff}(T)(Y_2 -Y_1^2\frac{\bar{Y}_2}{\bar{Y}_1^2})
\label{eq:Y2}
\end{eqnarray}

\noindent
where  $Y_{1,2}$ are the abundances of each DM, $\bar{Y}_{1,2}$, the corresponding equilibrium abundances, $s$ is the entropy density, $H$ the Hubble parameter,
$\langle v\sigma^{ijkl}\rangle$ the thermally averaged cross-sections for all processes of the type $i,j\rightarrow k,l$ involving particles in sectors  0,1 and 2. In this equation there is an implicit summation over all particles that are involved in a given subprocess, e.g. $\langle v \sigma^{2200}\rangle$ includes all processes involving the annihilation of pairs of $\tilde{H^+},\tilde{H},\tilde{A}$  into pairs of SM particles. In Eq.~\ref{eq:Y2}, the term in
$\Gamma_{eff}$  describes the freeze-in production of $\tilde{S}$ from the decay of scalars in sector 2, 
\begin{equation}
\Gamma_{eff}(T) = \frac
{ \sum \limits_{\tilde{h} \in \{ \tilde{A}, \tilde{H}\}} g_{\tilde{h}} m_{\tilde{h}}^2 K_1(\frac{m_{\tilde{h}}}{T}) \Gamma_{\tilde{h}\to \tilde{S},\tilde{S} } }
{ \sum \limits_{\tilde{h} \in \{ \tilde{A}, \tilde{H}\}} g_{\tilde{h}} m_{\tilde{h}}^2 K_2(\frac{m_{\tilde{h}}}{T})  }
\end{equation}
$\Gamma_{\tilde{h}\to \tilde{S}\tilde{S}}$ is the partial width for the decay of $H_i=\tilde{H},\tilde{A}$ into two singlets,  and $m_i$ are the corresponding masses. $K_i(x)$ are the Bessel functions of second order and degree $i$.

To solve these equations we use micrOMEGAs5.3~\cite{Belanger:2018mqt,micro_prep} with  two different methods. The first involves the direct solution of  Equations~\ref{eq:Y2} and makes use of the function  {\tt darkOmegaN}. To validate these results we compare them  with the standard micrOMEGAs5.3 routines, that is {\tt darkOmega2} which is designed for WIMPs and will give the correct result for $\Omega_2$ and {\tt darkOmegaFI} which computes the relic density of the FIMP. We found good agreement between the two methods when the doublet cannot decay into singlets, when decays are allowed, {\tt darkOmega2} cannot be used as it does not include decays.  

Note that the decay of SM particles into the FIMP are effectively  included through the s-channel resonance in the annihilation cross-section $v\sigma^{1100}$, for example production from Higgs decay is computed as $b\bar{b}\rightarrow h \rightarrow \tilde{S}\tilde{S}^\dagger$.  The initial conditions at $T_R$, the reheating temperature, are taken to be $Y_1=0$ for the FIMP and 
$Y_2=\bar{Y}_2$ for the WIMP. Here we assume that both particles follow   Maxwell-Boltzmann statistics. For the impact of the quantum statistical distribution in the case of freeze-in, see Ref~\cite{Belanger:2018ccd}.

\begin{figure}[htb]		
	\includegraphics[scale=0.3]{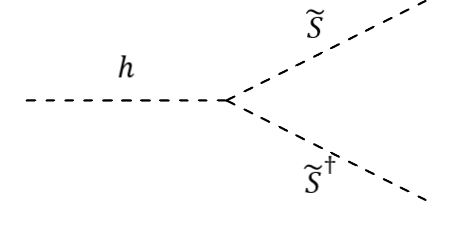}
	\includegraphics[scale=0.3]{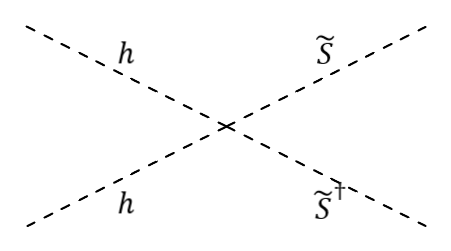}
	\includegraphics[scale=0.3]{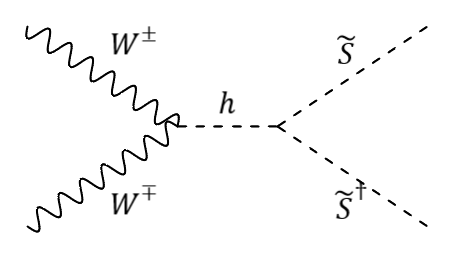}
	\includegraphics[scale=0.3]{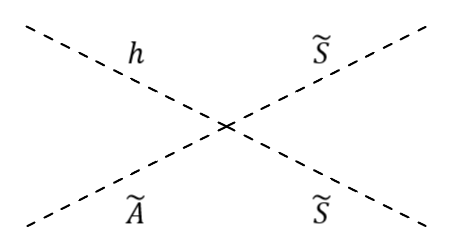}
	\includegraphics[scale=0.3]{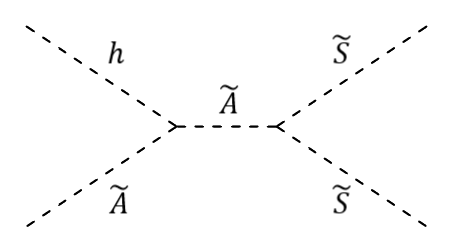}
		\includegraphics[scale=0.25]{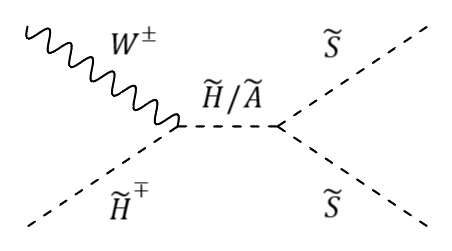}	
		\includegraphics[scale=0.3]{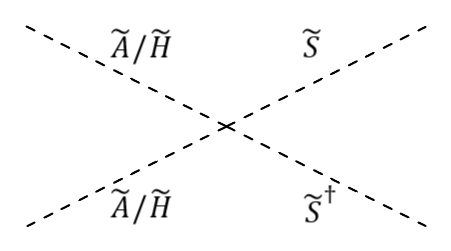}
		\includegraphics[scale=0.3]{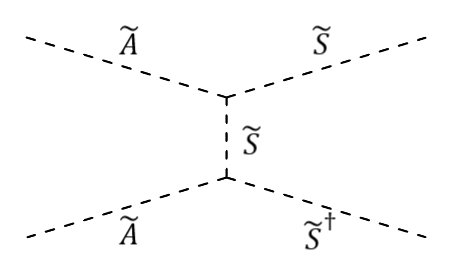}
		\includegraphics[scale=0.3]{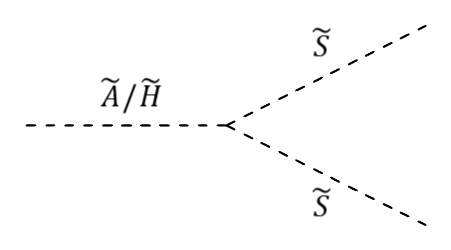}
	
	\centering
	\caption{Sample Feynman diagrams which contribute to singlet DM production. Top row: production from SM particles,  middle row : production from doublet and SM particles, bottom row : production from doublet components. }   
	\label{fig:diagrams} 
\end{figure}

\begin{figure}[h]
	\includegraphics[scale=0.5]{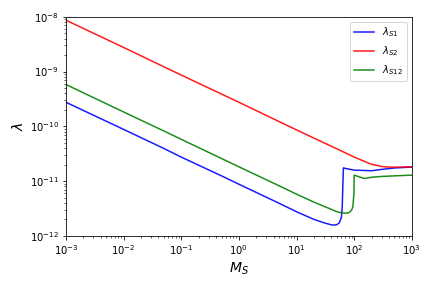}
	\includegraphics[scale=0.5]{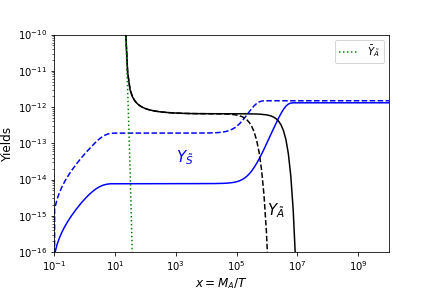}
	\centering
	\caption{Left panel: Contour curves corresponding to $\Omega_1h^2=0.12$ as function of $M_S$ for different simplified scenario where only one of the FIMP coupling is non-zero, either $\lambda_{S1}$(blue), $\lambda_{S2}$ (red) or $\lambda_{S21}=-\lambda_{S12}$ (green). Other parameters are fixed to  $M_A=198.6~{\rm GeV},M_H= 291.0~{\rm GeV}, M_{H^+}= 260.3~{\rm GeV},\lambda_3= 5.86$ and $\lambda_2= 4.41\times 10^{-3}$. Right panel : Abundances of $\tilde{S}$ and $\tilde{A}$  as function of $x=M_A/T$. The model parameters are fixed to 
	$M_S=363.1~{\rm GeV}$,$M_A=793.9~{\rm GeV},M_H= 795.1~{\rm GeV}, M_{H^+}= 799.7~{\rm GeV}$,$\lambda_2= 0.19,\lambda_3= -1.32\times 10^{-3}$, $\lambda_{S1}=-1.21\times 10^{-12},
	\lambda_{S2}=-7.44\times 10^{-15}$,$\lambda_{S12}=-2.34\times 10^{-13}$ and $\lambda_{S21}=-5.49\times 10^{-13}$ (full lines) or $\lambda_{S21}=-2.49\times 10^{-11}$ (dashed lines). The dotted line shows the equilibrium abundance of the doublet $\bar{Y}_{\tilde A}$. 
	}   
	\label{fig:omega1} 
\end{figure}

We first discuss the expected behaviour of the relic density under different simplified scenarios, general results will be presented in section ~\ref{sec:results}.   
The first example includes the case where all interactions between the doublet and singlet are negligible, $\lambda_{S2}=\lambda_{S12}=\lambda_{S21}=0$. Here freeze-in occurs through interactions of the singlet with the Higgs, mainly from Higgs decay if kinematically accessible, else from annihilation of a pair of SM particles through the s-channel exchange of a Higgs, see Fig.\ref{fig:diagrams}. The two dark sectors are decoupled and it is possible for $\tilde{S}$ to be the dominant DM component for any values of DM masses,   the value of  
$\lambda_{S1}$  leading  to $\Omega_1 h^2=0.12$ is shownFig.~\ref{fig:omega1}. For $M_S=m_h/2$, $\Omega_1 h^2 \propto \lambda_{S1}62/M_S$ so that the value of $\lambda_{S1}$ decreases with $M_S$.  When the decay channel is closed, a sharp increase in $\lambda_{S1}$ by roughly one order of magnitude is required to reach the same value of $\Omega_1h^2$. In this figure we fix the parameters of the doublet sector to
 $M_A=198.6~{\rm GeV},M_H= 291~{\rm GeV}, M_{H^+}= 260.3~{\rm GeV},\lambda_3= 5.86,\lambda_2= 4.41\times 10^{-3}$, which leads to $\Omega_2 h^2=3.4\times 10^{-6}$. 
In this scenario, the doublet behaves as in the IDM and can account for any fraction of DM although it can be dominant only when $M_A>550 {\rm GeV}$\cite{Belanger:2021lwd}. If the fraction of the doublet component DM  is small, it will more easily escape direct or indirect searches.  The singlet FIMP being basically undetectable, the only DM signatures will be found for the doublet component and  the phenomenology of this class of scenarios is similar to the one of the Inert Doublet model where the doublet is allowed to be a subdominant DM component. 

The second case is the one where $\lambda_{S1}=\lambda_{S12}=\lambda_{S21}=0$ such that there are no interactions between $\tilde{S}$ and the SM, instead freeze-in occurs through interactions with the inert doublet. Yet, decays of the doublet into a pair of singlets are not allowed at tree-level, the dominant process for freeze-in is then $\tilde{A}\tilde{A}  \rightarrow \tilde{S} \tilde{S}^\dagger$ through the quartic interaction, see Fig.~\ref{fig:diagrams}. The relic density $\Omega_1 h^2\propto \lambda_{S2}^2 M_S$ modulo some factors taking into account the  effective degrees of freedom. Although the singlet freezes-in through the second dark sector, as above  it can be the dominant DM component  for any values of $M_S$, see Fig. ~\ref{fig:omega1}.  Moreover, interactions between the singlet and doublet  have little effect on the relic density of the doublet considering the small couplings involved. As the previous case,  the collider and astroparticle phenomenology in  these scenarios are similar to the IDM where one allows a subdominant doublet DM. 

The third possibility is the one where only semi-annihilation processes play a role in freeze-in, that is  $\lambda_{S1}=\lambda_{S2}=0$, $\lambda_{S12},\lambda_{S21}\neq 0$. When  $M_{A} < 2M_{S}$,  this scenario still leads to two DM and $2\to 2$ processes are responsible for  freeze-in.  If $M_A+m_h >2M_S$,  the semi-annihilation  channel $h\tilde{A}\rightarrow \tilde{S}\tilde{S}$  is dominant, otherwise  processes such as  $\tilde{A} \tilde{A}(\tilde{H} \tilde{H}) \rightarrow \tilde{S}\tilde{S}^\dagger$ dominate the DM formation in the first sector, the relevant diagrams are displayed in Fig.\ref{fig:diagrams}. 
The value of the coupling  required for the singlet to form all of DM is shown in Fig.~\ref{fig:omega1} where we assume
$\lambda_{S12}=-\lambda_{S21}$ to maximise the couplings to the pseudoscalar doublet to the singlet. The coupling is highest for a very light scalar and decreases linearly when $M_S$ increases until $M_S>M_A/2$ where decay ceases.

When  $M_{A} > 2M_{S}$,  the decays $\tilde{A}, \tilde{H} \to \tilde{S} \tilde{S}$ become important for DM formation and one expects only one stable  DM component, $\tilde{S}$.
The main production for the singlet FIMP can occur at temperatures above $T_{FO}$, the freeze-out temperature of the doublet, this is freeze-in production. The doublet can also decay into singlets after $T_{FO}$, this is called the super-WIMP mechanism. The total relic density of the singlet can be split into two components
\begin{equation}
\Omega_1=\Omega_1^{FI}+\Omega_1^{SWIMP}   \;\; {\rm where} \;\;\Omega_1^{SWIMP}=\frac{2M_S}{M_A} \Omega_A^{FO}
\label{eq:FI}
\end{equation}
When computing the relic density, we have solved the two coupled abundance equations (Eq.~\ref{eq:Y2}) until $T\approx 10^{-8}${\rm GeV}. We  have also checked that the resulting value of $\Omega_1$  matched with  the one obtained by solving independently for $\Omega_1^{FI}$ and $\Omega_A^{FO}$ and applying Eq.~\ref{eq:FI}.

The evolution of the abundances with temperature is illustrated  for two typical points in Fig.~\ref{fig:omega1} (right).  At high temperatures,  $\tilde{S}$ is produced mainly from decays of $\tilde{H},\tilde{A}$ with some subdominant contribution from $W^+,\tilde{H^-}\rightarrow \tilde{S},\tilde{S}$. Freeze-in production stops around $x=M_A/T=3$. At that point the doublet is still in thermal equilibrium with the bath. Freeze-out of the doublet occurs near $x=20$. At this time, DM is mostly composed of the lightest doublet with $Y_{\tilde{A}} \gg Y_{\tilde{S}}$ as can be seen in Fig.~\ref{fig:omega1}. The doublet then decays into  singlets which then  form all of the DM today.   
For larger couplings between the doublet and the singlet, a larger fraction of DM would be formed by freeze-in  before the doublets freezes-out. This is illustrated in 
Fig.~\ref{fig:omega1} (right) where the dashed lines  correspond to a larger value of $\lambda_{S21}=-2.49\times 10^{-11}$.
Scenarios with one DM will be discussed in detail in Section~\ref{sec:oneDM}.

\subsection{Direct and Indirect detection}
\label{sec:ID}
The signals from FIMP dark matter are very much suppressed in both direct and indirect searches considering the small couplings involved. While in the case of  very light mediators, it was shown that the direct detection cross-section could be detectable for a FIMP, in the Z4IDSM  model the Higgs is the mediator, thus the direct detection cross-section for the FIMP is negligible. The DM signal therefore could arise only for the WIMP component. The signal for direct detection  is suppressed by the fraction of the second dark matter, $\xi_2$ where 
\begin{equation}
\xi_i=\Omega_i h^2/(\Omega_1  h^2+\Omega_2 h^2).
\end{equation}
 The quantity that can be directly compared with the limit from direct detection experiments for spin-independent interactions is the  rescaled  DM-proton cross-section, $\sigma_{Ap} \xi_2$ where $\sigma_{Ap}$ is computed with micrOMEGAS~\cite{Belanger:2008sj,Belanger:2020gnr} assuming a local DM density $\rho=0.3 g/cm^2$ and a Maxwellian velocity distribution.

Since only the doublet DM will contribute to the indirect detection cross-section, we can write the total number of events per DM pair annihilation as
 \begin{equation}
 N=  \rho^2 \frac{\xi_2^2}{2 M_A^2}  \langle v\sigma^{AA} \rangle 
 \end{equation}
 where $\langle v\sigma^{AA}\rangle$ is  the cross-section for annihilation of $\tilde{A}\tilde{A}$ into all SM final states. It is related to the total cross section for annihilations into all SM final states, $\langle v \sigma \rangle$
 \begin{equation}
\langle v\sigma^{AA}\rangle \left(\frac{\xi_2}{M_A}\right)^2 = \langle v \sigma \rangle  \left(\frac{\xi_1}{M_S}+\frac{\xi_2}{M_A}\right)^2
\end{equation}
The production rate of particles $a$ (here photons or antiprotons) from DM annihilation, can be written as 
\begin{eqnarray}
 Q^a(E) &=&  \frac{1}{2} \rho^2 \left(\frac{\xi_2}{M_A}\right)^2\langle v\sigma\rangle \frac{dN^a_{AA}(E)}{dE}
\end{eqnarray}
and $dN^a_{AA}/dE$ is the energy distribution of particle $a$ produced in $\tilde{A}\tilde{A}$ annihilations.

\section{Constraints}
	\label{sec:constraints}
We apply several theoretical, collider and astrophysical constraints on the model.\\ 	

\noindent
{\bf Theoretical constraints}\\

We impose a perturbativity condition that the vertex factor in the Feynman rules for quartic interactions  must be smaller than $4\pi$. This is  to ensure  that the one-loop corrections are smaller  than the tree-level contributions ~\cite{Lerner:2009xg}. This leads to the conditions, 
  \begin{eqnarray}
  &&\lambda_2<\frac{2\pi}{3} ,\;   |\lambda_3| < 4\pi ,\;  |\lambda_3+\lambda_4| < 4\pi ,\;  |\lambda_4\pm\lambda_5| < 8\pi ,\;  |\lambda_3+\lambda_4\pm \lambda_5| < 4\pi ,\;  \nonumber\\
   && |\lambda_5| < 2\pi \;,\;\;  |\lambda_S|<\pi  \,,\;\;     |\lambda'_S|<\frac{\pi}{3} \;. 
  \end{eqnarray}
  while $|\lambda_{S1}|, |\lambda_{S2}|<4\pi$ are always satisfied in the singlet FIMP scenario.
   Moreover we require that the  partial wave unitarity condition should be satisfied for all scattering amplitudes as described in ~\cite{Belanger:2014bga} and that the potential is bounded from below leading to the stability conditions on the potential  listed in  ~\cite{Belanger:2014bga}. In addition, we impose the condition that the  inert vacuum corresponds to the global minimum of the potential. Since the couplings of the singlet are very small, we recover approximately the potential of the inert doublet and impose the conditions on the global minimum of that potential, see~\cite{Ginzburg:2010wa}.   We introduce $R=\lambda_{Ah}/2\sqrt{\lambda_1 \lambda_2}$ which satisfies $R>-1$, and we require, apart from $\lambda_1>0$, the following conditions on the parameter of the potential,
\begin{equation}
\mu_2^2> -R\sqrt{\lambda_1\lambda_2} v^2, \hspace{5pt} {\rm if} \;\;  |R|<1,
\nonumber
\end{equation}
\begin{equation}
\mu_2^2>-\sqrt{\lambda_1\lambda_2} v^2, \hspace{5pt} {\rm if} \;\; R>1. 
\end{equation}

\noindent
{\bf Collider constraints}\\

The electroweak precision parameters S,T are sensitive to physics beyond the SM. In this model the one-loop contribution arises solely from the new scalar  doublet and are similar to the IDM. Assuming a SM Higgs boson mass  $m_{h}=125$ GeV,  the  values of the S and T parameters in the limit U=0, have been determined to be 
\begin{equation}
S=0.06\pm0.09, \hspace{12pt} T=0.1\pm 0.07
\end{equation}
with a correlation coefficient +0.91\cite{Baak:2014ora}.   When scanning over the parameter space we have computed S and T for each point and have required that they  fall within the ellipse corresponding to the 95\% C.L. limit.

In the low mass range, LEP experiments constrain the parameter space of the model, these limits are similar to the ones obtained in the IDM, and include those obtained
from the  measurements of the widths of the $W$ and $Z$ bosons,
\begin{equation}
M_A + M_{H^+}> M_W\;,\;\;\; M_H +  M_{H^+} > M_W\;,\;\;\;
M_H+ M_A >M_Z \;,\;\;\; 2 M_{H^+} > M_Z
\end{equation}
as well as limits from  searches for $e^+e^- \rightarrow \tilde{H}\tilde{A}$~\cite{Lundstrom:2008ai}. 
    \begin{equation}
    M_H<80 {\rm GeV} \;,\;\;M_A <100 {\rm GeV}\;,\;\; {\rm if} \;\;\; M_A - M_H>8 {\rm GeV}
    \end{equation} 
   A lower limit on the charged Higgs mass,   
 $M_{H^+}>70 GeV$,
results from a re-interpretation of limits on charginos \cite{Pierce:2007ut}. This limit increases to $M_{H^+}>100$GeV~\cite{OPAL:2003zpa} if  the  charged Higgs is long-lived with a lifetime $\tau>10^{-6}$ sec. The
 theoretical constraints (stability and unitarity) as well as  Electroweak Precision Test limits and LEP constraints, lead to an upper limit on the mass splitting within the scalar doublet~\cite{Belanger:2021lwd}.

The LHC can further constrain the model.  Precise measurements of the properties of the Higgs allow to put an upper limit on the invisible decay of the Higgs into dark matter, 
    	$Br\left( h \rightarrow invisible\right)< 11\%$~\cite{ATLAS_invisible}. The  loop-induced partial decay width for $h \rightarrow \gamma \gamma$  receives a contribution from the charged Higgs, the partial decay width predicted  in the Z4IDSM  is compared with the experimental upper bound using HiggsSignals~\cite{Bechtle:2020uwn}. 

Finally new physics searches at the LHC allow to probe the model, those are monojet searches, disappearing track searches, searches for heavy stable charged particles (HSCP) and dileptons. These constraints apply to the doublet sector and are thus similar to those found in the IDM. For the  monojet search we use a recast   performed in ~\cite{Belyaev:2018ext} and ~\cite{Belyaev:2016lok} in the IDM. We will use the exclusion for 20 - 100 fb$^{-1}$ as well as the projection for 3ab$^{-1}$ and will show that they have little impact on the parameter space of the Z4IDSM model.  When the mass splitting between the charged Higgs and DM is small (at most few hundred MeV's), the charged Higgs is long-lived. There are two collider searches aimed at long-lived particles  that can be relevant in this model : the search for heavy charged particles (HSCP) and the disappearing track search which correspond to the charged Higgs decaying into soft particles and DM with $c\tau\approx m$. For these searches we use the limits implemented via SModelS2.0.0~\cite{Heisig:2018kfq,Ambrogi:2018ujg,Khosa:2020zar,Alguero:2020grj}, where  the HSCP limits  are  based on ~\cite{CMS:2015lsu,ATLAS:2019gqq}. For disappearing tracks our results are consistent with~\cite{Belyaev:2020wok}. Finally we ignore dilepton constraints as it was shown in the IDM that  dilepton searches constrain mostly the region where the doublet is below the electroweak scale~\cite{Belanger:2015kga,Chakraborti:2018aae}, a region that is already severely constrained. \\

\noindent
{\bf Dark matter constraints}\\

We  require that the total relic density falls within the observed range determined
very precisely by the PLANCK collaboration, to $\Omega h^2= 0.1184 \pm 0.0012$~\cite{Planck:2015fie}. We also take into account a 10\% theoretical uncertainty, and take the $2\sigma$ range, which corresponds to 
\begin{equation}
0.094<\Omega_{DM} h^2 < 0.142.
\end{equation}

 Scalar DM interacts with nuclei only through spin-independent (SI) interactions. For DM masses near  the electroweak scale,  as considered  here we use the  limits on the SI cross-section for DM scattering on nucleons  obtained  by XENON1T~\cite{Aprile:2018dbl} . We compute the recoil energy distribution corresponding to the doublet  scaled by their fractional contribution to the total DM density,  and use the recasted limits at 90\%C.L.  implemented in \micro~\cite{Belanger:2020gnr}. For this we use the default values for the astrophysical parameters, namely a local DM density $\rho_{DM}= 0.3 {\rm GeV/cm}^3$, and a standard Maxwellian velocity distribution.
We also show projections for XENON-nT~\cite{XENON:2015gkh} and DARWIN~\cite{DARWIN:2016hyl}, note that recent constraints  from PandaX~\cite{PandaX-II:2020oim} will not be displayed explicitly as they lie between the current and projected XENON's results.

Fermi-LAT observations of photons from Dwarf Spheroidal Galaxies (dSph), provide one of the most robust constraint on DM. In our model, as will be discussed later, the DM annihilation cross-sections  are in general largest for the gauge bosons final states.  Thus we compute $ \langle v \sigma \rangle_{VV}= \langle v\sigma^{AAWW} \rangle +\langle v\sigma^{AAZZ} \rangle$ corresponding to the annihilation of $AA$ into  WW and ZZ final states  and require that  $\langle v \sigma \rangle_{VV}$
 lies below the 95\%C.L. given by FermiLAT in ~\cite{Fermi-LAT:2015att}.  We ignore the small  difference between the photon spectra from W and Z's. This approach is conservative as other channels can contribute as well. We also impose constraints from anti-proton searches
 performed by AMS-02~\cite{AMS:2016oqu}. We use  the limits derived in ~\cite{Reinert:2017aga} for WW final states, and apply it to both the WW and ZZ final states,  the antiproton spectrum from WW and ZZ being close to each other. These limits take into account the uncertainties on the cosmic ray propagation parameters and  are derived from  a global fit to B/C ~\cite{AMS:2016brs} and to the antiproton spectrum. We will use the most conservative limit obtained assuming a generalised NFW profile with $\rho=0.3 {\rm GeV/cm}^3$ and also display the limit for a standard NFW profile with $\rho=0.38 {\rm GeV/cm}^3$.

Limits on DM annihilation  will also be obtained by CTA  which  measures the photon spectrum albeit at higher energies~\cite{Acharyya:2020sbj}. We perform a dedicated analysis to determine the parameter space of the model that is  within reach of CTA. For this we use the combined photon spectra from all annihilation channels.\\

\noindent
{\bf Big-Bang Nucleosynthesis}\\
\label{sec:BBN}

When the decay of the doublet into singlets is kinematically allowed, the decays $\tilde{A} \rightarrow Z^{(*)} \tilde{S}\tilde{S}$ or $\tilde{H^+} \rightarrow W^{(*)} \tilde{S}\tilde{S}$ can be long-lived and could disrupt the successful predictions of BBN. In particular, injection of hadronic energy from decays with lifetime around 100s,  can 
trigger non-thermal nuclear processes and  affect the light element adundances. Constraints on the amount of hadronic energy released in the late decay of the WIMP after it freezes-out were derived in ~\cite{Kawasaki:2004qu}, we follow this approach and include the newer determination of $^2H$ and $^4He$~\cite{Iocco:2008va,Cyburt:2015mya} as in \cite{Banerjee:2016uyt}.
The hadronic energy released is defined as
\begin{equation}
\xi_{had}= \epsilon_{had} B_{had} Y_{2}
\end{equation}
where $B_{had}$ is the branching fraction of $\tilde{A}$ or $\tilde{H^+}$ into hadronic components, roughly it is given by
\begin{eqnarray}
  B_{had}(\tilde{A}) &=&\frac{ \Gamma (\tilde{A} \rightarrow Z^{(*)} \tilde{S}\tilde{S})B_{had}(Z)+ \Gamma (\tilde{A} \rightarrow h^{(*)} \tilde{S}\tilde{S})B_{had}(h)}{ \Gamma_{\rm TOT} (\tilde{A})}
  \end{eqnarray}
  for the neutral component while for $\tilde{H}^+$ which can only decay  through real or virtual W's, we use
\begin{equation}  
 B_{had}(\tilde{H}^+)=B_{had}(W^+)
\end{equation}
Note that  the decay through a virtual h depends on the same parameters as the two-body decay ($\lambda_{S12}-\lambda_{S21})$, and is therefore always suppressed. On the other hand the three-body decay through a Z can become dominant in the limit $\lambda_{S12}\approx \lambda_{S21}$.
$\epsilon_{had}$ is the hadronic energy released in each WIMP decay while $Y_{2}$ is the WIMP abundance after it freezes-out. This quantity is computed with micrOMEGAs as described in section~\ref{sec:relic}. For decays through a real W or Z the hadronic energy is taken to be
\begin{equation}
\epsilon_{had}\approx \frac{m_{A,H^+}^2- (2m_S)^2 +m^2_{Z,W}}{2m_{A,H^+}}
\end{equation}
where we treat the two singlets as a single particle,  this corresponds roughly to the peak of the hadronic energy distribution. 
For 4-body decays, $\tilde{A}\rightarrow q\bar{q} \tilde{S}\tilde{S}$ through a virtual Z or  $\tilde{H^+}\rightarrow q\bar{q'} \tilde{S}\tilde{S}$ through a virtual W, we use
\begin{equation}
\epsilon_{had}\approx \frac{m_{A,H^+}^2- (2m_S)^2 }{2m_{A,H^+}}
\end{equation}
We follow the approach in ~\cite{Banerjee:2016uyt} which is based on the results derived in ~\cite{Kawasaki:2004qu} but take into account newer determinations of the abundance of light elements, for reviews see ~\cite{Iocco:2008va,Cyburt:2015mya}. \\

\noindent
{\bf CMB}
\label{sec:CMB}

Long-lived particles can impact the CMB leading to spectral distortions  as well as to anisotropies in the power spectra.
The former provide constraints of the amount of energy injected at early epochs (before $10^{12}$ sec)  while CMB anisotropies constrain mostly energy injection from decaying particles  with $\tau> 10^{12}$sec.  To estimate the effect of long-lived particles on the CMB, we follow the analysis in ~\cite{Poulin:2016anj}  where the effect of electromagnetic decays on the CMB power spectra  and the corresponding PLANCK constraints were given in terms of the effective energy density, $\Xi$, and the lifetime of the decaying particle where 
\begin{equation}
\Xi= \frac{\Omega_2}{\Omega_1+\Omega_2} E_{em}
\end{equation}
and $E_{em}$ is the fraction of the decay energy taken by photons and $e^\pm$. Moreover  in Ref.~\cite{Poulin:2016anj}, spectral distorsion constraints from FIRAS~\cite{Fixsen:1996nj} are also expressed in terms of the effective energy density. These constraints are relevant in specific scenarios where  one component of the  doublet can decay  into pairs of DM singlets with a very long lifetime as will be discussed in section~\ref{sec:CMB}.

\section{Results}
\label{sec:results}

After having presented the simplest case where only one freeze-in coupling is non-zero, we investigate the full parameter space of the model. For this we perform a random scan over the free parameters of the model as shown in Table~\ref{tab:range}. We first concentrate on the choice $M_A<M_S/2$ in order to guarantee two DM candidates for any choice of couplings. 
The case  of the light singlet will be discussed in section~\ref{sec:oneDM}.

 \begin{table}[!htb]
\begin{center}
\begin{tabular}{|cc|cc|cc|}
\hline
     $M_S$   &$40 - 1500$ GeV &   $\lambda_2$&$10^{-5} - 2\pi/3$ &$|\lambda_{S1}|$ & $10^{-16} - 10^{-10}$ \\
 $M_A$ & $40 - 1000$   GeV&   $|\lambda_3|$&$10^{-5} - 4\pi$&$|\lambda_{S2}|$      & $10^{-16} - 10^{-10}$ \\
    $\Delta^+$  &$0 - 500 $ GeV&   $\lambda_S$&$1$&$|\lambda_{S12}|$   & $10^{-16} - 10^{-10}$ \\
  $\Delta^0$  &$0 - 500 $GeV& $|\lambda'_S|$&$1/2$&$|\lambda_{S21}|$   & $10^{-16} - 10^{-10}$ \\
     \hline
     \end{tabular}
     \caption{Range of the free parameters of the Z4IDSM model used in the scan.}
     \label{tab:range}
     \end{center}
\end{table}

\subsection{Relic density and direct detection}

\begin{figure}[h]
	\includegraphics[scale=0.5]{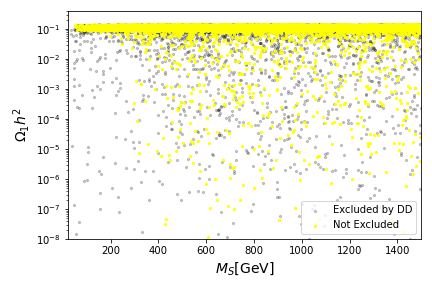}
	\includegraphics[scale=0.5]{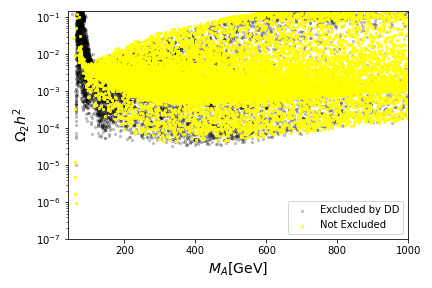}
	\centering
	\caption{$\Omega_1h^2$ (left)  and  $\Omega_2h^2$ (right) as a function of the corresponding  DM mass for all points satisfying theoretical, collider and the total relic density constraints (grey) as well as DD constraint from XENON-1T(yellow). }   
	\label{fig:omega} 
\end{figure}

 As expected from the discussion in Section~\ref{sec:relic} the singlet FIMP dark matter can account for most of the DM in the universe  as long as $M_S>53~{\rm GeV}$, see Fig.~\ref{fig:omega} (left). 
 The doublet will typically be a subdominant component, although the doublet can form most of the DM when its mass is near $m_h/2$ or above  550~GeV, as expected from the inert doublet scenario.  For the range of parameters in our scan  $\Omega_2 h^2$ can take any value above a few  $10^{-5}$ except when $M_A \approx m_h/2$ where it can be lower, see Fig.~\ref{fig:omega} (right). The lowest value of $\Omega_2 h^2$  for each $M_A$  is determined by the  largest allowed value for  $\lambda_{Ah}$ which is basically determined by theoretical constraints on the potential.

\begin{figure}[htb]
	\includegraphics[scale=0.5]{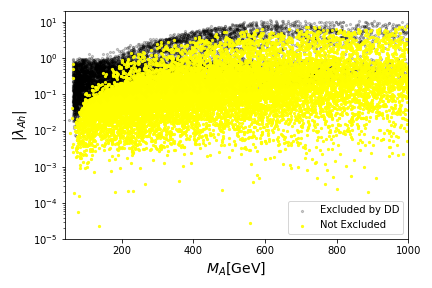}
	\includegraphics[scale=0.5]{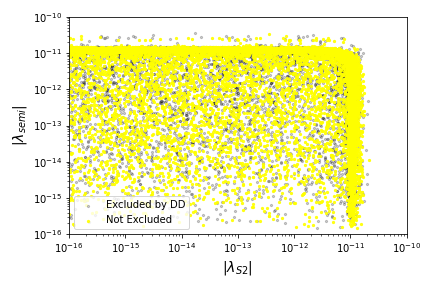}
	\centering
	\caption{Left: Allowed points in the $\lambda_{Ah} -M_A$ plane (left)  and  $\lambda_{semi}$ - $\lambda_{S2}$ plane  (right)  same color code as Fig.~\ref{fig:omega}. }   
	\label{fig:semi} 
\end{figure}

  In this model, the SI cross-section for the FIMP is extremely suppressed thus DD constraints apply only for the doublet component.  After direct detection constraints are imposed only a small fraction of the region $M_A< m_h/2$ remains possible with $M_A>55.9~{\rm GeV}$. 
 On the other hand the full range of values of $\Omega_2 h^2$ satisfies DD constraints for heavier $M_A$. Indeed the relic density, when dominated by DM pair annihilation typically goes as $1/\lambda_{Ah}^2$ corresponding to annihilation into WW,ZZ,hh,$f\bar{f}$ through a diagram with exchange of $h$ in the s-channel. The cross-section for direct detection is proportional to the same coupling $\lambda_{Ah}^2$, thus the rescaled SI cross-section $\sigma^{SI} \xi_2$ is roughly constant. Since the DD limit weakens at higher masses, more points are allowed for heavier DM.   Moreover at large DM masses the annihilation into gauge boson pairs receives an additional contribution from the quartic diagram thus loosening the connection between the relic density and direct detection.  In some cases the relic density is not dominated by the s-channel annihilation process, for example when $m_A>m_h$ there is also a contribution to DM pair annihilation into hh from a t-channel diagram, this contribution which goes as $\lambda_{Ah}^4$ can become dominant at large values of $\lambda_{Ah}>1$ in this case the rescaled SI cross-section $\sigma^{SI} \xi_2 \propto 1/\lambda_{Ah}^2$. Hence larger values of $\lambda_{Ah}$ are allowed for large $M_A$ as can be seen in Fig.~\ref{fig:semi} (left) which shows that the limit
 from XENON-1T leads to an upper bound for $\lambda_{Ah}$ which increases with $M_A$.
  Other couplings cover the full range used in the scan and are not directly subjected to DD constraints. A large fraction of the points that reach the correct relic density are concentrated in the region $\lambda_{semi}=\sqrt{\lambda_{S21}^2+ \lambda_{S12}^2} \approx 10^{-11}$ or $\lambda_{S2} \approx 10^{-11}$ or $\lambda_{S1} \approx 10^{-11}$. The distribution of the allowed points in the $|\lambda_{semi}| - |\lambda_{S2}|$ plane is shown in Fig.~\ref{fig:semi} (right).
  
\begin{figure}[hbt]
	\includegraphics[scale=0.45]{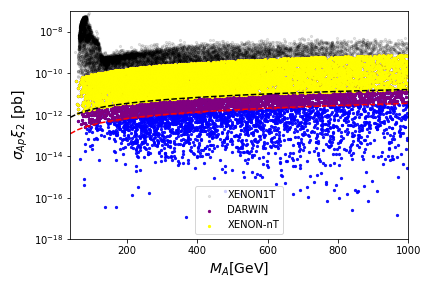}
	\centering
	\caption{ Spin-independent DM proton scattering cross section   times the fraction of the second DM component, $\sigma_{Ap} \xi_2$  as  function of $M_A$. 
	Points in black are ruled out by XENON-1T, points in yellow (red) are within the reach of XENON-nT (DARWIN) and blue points are beyond the reach of DARWIN. The black (red) dotted line shows the projected reach of XENON-nT (DARWIN), }   
	\label{fig:sigma} 
\end{figure}

Future multi-ton detectors such as XENON-nT and DARWIN offer the possibility to further probe the model, covering a large fraction of the currently allowed parameter space, see Fig.~\ref{fig:sigma}. However, in some scenarios the signal is suppressed  and can be up to four order of magnitudes below the expected reach of DARWIN, this is below the neutrino floor ( Fig.~\ref{fig:sigma}) and therefore beyond the reach of even larger detectors.

\subsection{Colliders, Direct and Indirect detection}
\label{sec:sub:ID}

As mentioned above, the only indirect detection signature comes from the doublet. Annihilation channels  into vector bosons are dominant. 
The largest cross-sections are found when $M_A>500$GeV and the doublet constitute a large fraction of the dark matter. This region is constrained mildly by FermiLAT limits from dSPhs and also by the AMS02 limit on antiprotons especially if one chooses a generalised NFW profile with a local density $\rho=0.38 {\rm GeV/cm}^3$. The prospects from CTA allow to cover most of the region where $\langle v \sigma\rangle_{WW} > 2. 10^{-26} {\rm cm}^3/{\rm s}$. Although most of the parameter space remains out of reach of ID.

 	\begin{figure}[h]
	\includegraphics[scale=0.45]{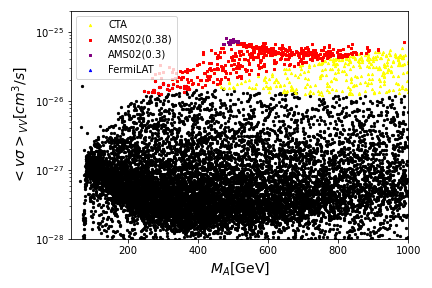}
	\includegraphics[scale=0.45]{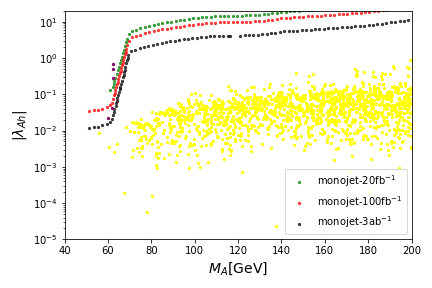}
	\centering
	\caption{ Constraints on $\langle v \sigma\rangle_{VV}$   from AMS02 searches on antiprotons  assuming $\rho=0.3$ (purple) or $\rho=0.38$ (red) and from FermiLAT (blue). Points in yellow indicate the future reach of CTA while points in black are out of reach of indirect searches.   	Monojet limit at LHC for ${\cal L}=20,100,3000 {\rm fb}^{-1}$ and allowed points  for the Z4IDSM model (blue) in the $|\lambda_{Ah}| - M_A$ and $|\lambda_{S1}| - M_S$ plane }   
	\label{fig:indirect} 
\end{figure}

\begin{figure}[h]
	\includegraphics[scale=0.5]{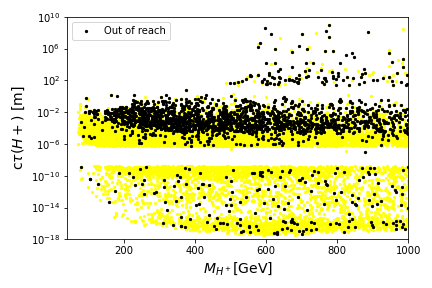}	
	\includegraphics[scale=0.5]{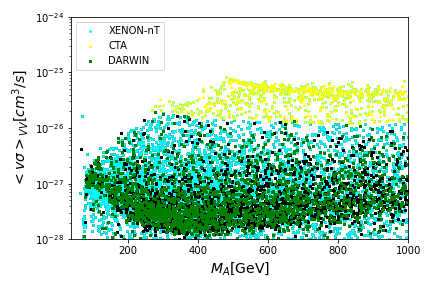}		
	\caption{Lifetime of the charged Higgs for currently allowed points (yellow) and for those that will be out of reach of future direct and indirect detection searches. 
	Future reach of direct detection with XENON-nT (cyan) and  DARWIN (green) and indirect detection with photons (CTA) in the $\langle v\sigma\rangle_{VV}-M_A$ plane (right panel). Points in black are beyond the reach of these detectors.}   
	\label{fig:futurereach} 
\end{figure}

 At the LHC the main signature of DM is through monojet,  $pp\rightarrow j \tilde{A}\tilde{A}$. This process allows  to probe  large values of $\lambda_{Ah}$, which are found for $M_A\approx 60{\rm GeV}$. 
 Searches for long-lived particles and in particular HSCP are more effective in constraining the model. Such searches currently rule out long-lived particles with masses below roughly 550 GeV.  The prediction for the life-time of the charged Higgs for all points currently allowed is displayed in  Fig.~\ref{fig:futurereach} (left), this   shows that only few of the points featuring  a HSCP signature with $c\tau > 1 {\rm m}$ remain at large values of $M_{H^+}$ while a major fraction of the parameter space falls within the displaced signature lifetime region $c\tau \approx 10^{-4} - 1 {\rm m}$ or within the prompt region.

The possibility to further probe the model in direct and indirect detection with future detectors such as XENON-nT, DARWIN or CTA is shown in the $\langle v\sigma \rangle_{VV}$- $M_A$ plane in Fig.~\ref{fig:futurereach} (right), clearly only the largest cross-sections will be probed by CTA, while XENON-nT and DARWIN can not only probe  the same region but
 also significantly extent the coverage of the parameter space. 
Isolating only scenarios that are beyond the reach of future direct and indirect searches, it becomes clear that exploiting the disappearing tracks signature at colliders  offer the best potential to further probe the model, this is illustrated in Fig.~\ref{fig:futurereach} (left) where we have selected in black the points that escape future direct or indirect searches. 
A detailed analysis is left for future work. 

\section{Results: one dark matter}
\label{sec:oneDM}

Here we discuss the case where the lightest component of the doublet can decay into the singlet, reducing effectively to a single dark matter model. We treat separately two scenarios, the one where the lightest component of the doublet is neutral and the one where it is charged. 

\subsection{The case where $M_A <M_{H+},M_H$}

For this scenario, we performed a dedicated scan choosing for  the  range of values for the feeble couplings  $10^{-15} - 10^{-9}$  while  the masses are varied in the range
\begin{equation}
M_S= 1-500 {\rm GeV}, \;\;\; 2 M_S <M_A < 1000  {\rm GeV}\; ,\;\;  \Delta^+,\Delta^0 < 500 {\rm GeV}.
\end{equation} 
Note that this scenario can  also accommodate singlets with masses below the GeV scale, for this larger values of the feeble couplings must be considered, see Fig.~\ref{fig:omega1}, hence we fix the upper limit of the feeble couplings to $10^{-9}$.   Other couplings are chosen as  in Table~\ref{tab:range}. 
As discussed in Section~\ref{sec:relic}, the decay of the doublet into DM ($\tilde{A}\rightarrow \tilde{S}\tilde{S}$) can occur before or after $\tilde{A}$ freezes-out. 
 We solved Eq.~\ref{eq:Y2}  to compute the abundance  of both components  at $T=10^{-8} {\rm GeV}$ and  we found that in all cases the doublet has completely decayed and  $\Omega_2 h^2\approx 0$. 
 
 The relic density constraint is satisfied for a few scenarios: 1) the FIMP singlet freezes-in mainly through its interactions with the SM sector while  the abundance of the doublet when it freezes-out is small and does not give additional contribution to $\Omega_1 h^2$;
2) the FIMP singlet  freezes-in through its interactions with the inert doublet  and might in addition receive a contribution from the doublet decaying after its FO, as discussed in section ~\ref{sec:relic};
 3) freeze-in occurs through interactions with both the SM and the inert doublet.  
The first scenario corresponds to values of $\lambda_{S1}$ ranging from $10^{-12}-10^{-10}$ according to the singlet mass, as shown in Fig.~\ref{fig:omega1}, 
and to small values of $\lambda_{S21},\lambda_{S12}$ and  $\lambda_{S2}$. Moreover a low abundance for the inert doublet at FO  is easily achieved  for $M_A<500$~GeV and/or large values of $\lambda_{Ah}$ as in the IDM. In the second scenario, small values of $\lambda_{S1}< 10^{-12}$ are possible and DM production is achieved  through interactions with the doublet. Thus it requires a non-negligible value for $\lambda_{S2}> 10^{-11}$ when production is dominated by freeze-in and/or for $\lambda_{S21},\lambda_{S12}$ when decay processes (before or after FO of the doublet) are important. Note that an important contribution from decays after FO requires that $\Omega_A^{FO}$ is not too small, hence is generally associated with a heavy doublet. The third scenario involves combinations of the previous two.

We find that the relic density constraint allows to cover the full range of masses considered for the singlet and the doublet. The configurations under consideration allow to avoid most
 constraints from DM observables. Indeed the abundance of the WIMP is negligible leading to strongly suppressed signatures in DD or ID   while the FIMP has no signature because of the small couplings. Thus, constraints on the model arise mainly from BBN and from colliders. 

\begin{figure}[h]
	\includegraphics[scale=0.5]{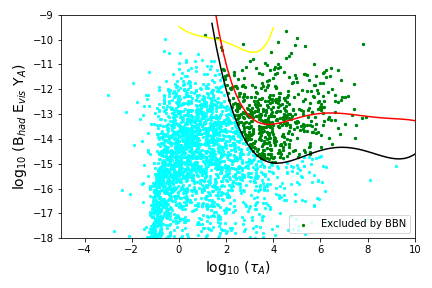}	
		\includegraphics[scale=0.5]{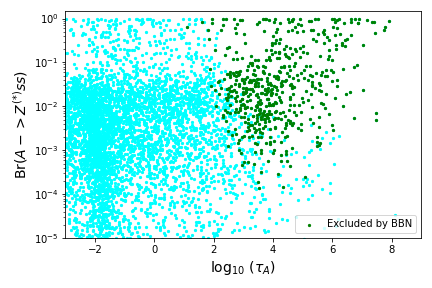}	
	\caption{Left: Allowed points in the injected hadronic energy vs lifetime plane  (cyan) and those  constrained by BBN (green). The curves
	show  the constraints from $^4He$ (yellow) and from $^2H/H$ (red) abundances. 
The black curve represents the  constraint from  assuming a tighter $^2H/H$ determination.
	Right: Branching fraction for  3-body and 4-body decays of $\tilde{A}\rightarrow Z^{(*)} \tilde{S}\tilde{S}$}   
	\label{fig:BBN} 
\end{figure}

\begin{figure}[h]
	\includegraphics[scale=0.5]{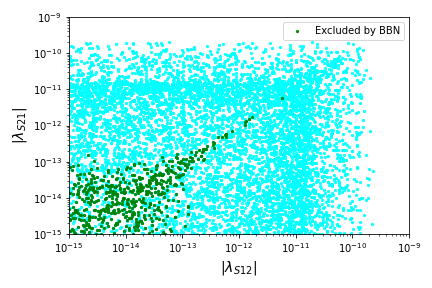}	
			\includegraphics[scale=0.5]{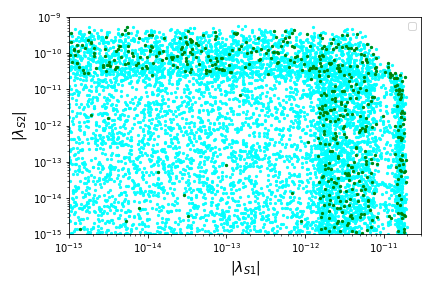}	
	\caption{ Left : Points that satisfy theoretical, collider and relic density constraints (cyan),  and those excluded by BBN (green) in the  $\lambda_{S21}$ vs $\lambda_{S12}$ plane (left) and  $\lambda_{S2}$ vs $\lambda_{S1}$ plane (right).
	}   
	\label{fig:param_A} 
\end{figure}

\begin{figure}[h]	
	\includegraphics[scale=0.5]{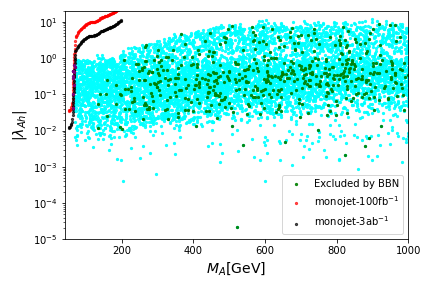}	
	\includegraphics[scale=0.5]{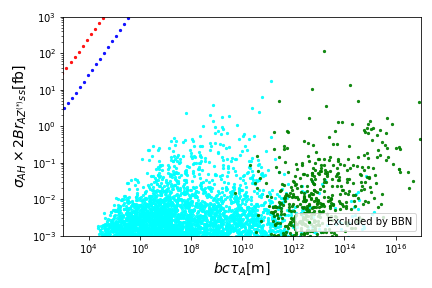}
	\caption{ Left :   Allowed points in the $\lambda_{Ah} - M_A$ plane, the monojet constraint (projection)  at the LHC for 100 $fb^{-1}$ and 3 $ab^{-1}$ is also shown. Same color code as Fig.~\ref{fig:param_A}.
	Right:  Projection for $\sigma(pp\rightarrow \tilde{A}\tilde{H})\times 2 Br(\tilde{A}\rightarrow Z\tilde{S}\tilde{S}$  vs $bc\tau_{A}$, the dotted lines show the reach of Mathusla from Ref.~\cite{Curtin:2018mvb}.   }
	\label{fig:LHC_A} 
\end{figure}

 BBN constraints are relevant  when  a particle has a long life-time  and decays through a hadronic channel. 
In our scan,  the decay $\tilde{A}\rightarrow \tilde{S}\tilde{S}$ is usually dominant and therefore has no impact on BBN, however in some cases  a significant fraction of $\tilde{A}$ decays through real or virtual Z ( $\tilde{A}\rightarrow Z^{(
*)}\tilde{S}\tilde{S}$) and are potentially constrained by  BBN when the lifetime of $\tilde{A}$ exceeds $100s$.  As described in Sec.~\ref{sec:BBN}, we compute $B_{had} \epsilon_{had} Y_A$  for each point in our scan, the results  are displayed in Figure~\ref{fig:BBN} (left).  Note that  a large fraction of the points in our scan predict $B_{had} \epsilon_{had} Y_A< 10^{-18}$ and are not shown in this figure. Figure~\ref{fig:BBN}  (right) shows that $Br(\tilde{A}\rightarrow Z\tilde{S}\tilde{S})$ must exceed $10^{-4}$ for a point to be excluded in addition to the condition $\tau_A>100$s. To determine excluded points we use the tighter constraint, it is clear from Fig.~\ref{fig:BBN} that uncertainties on the determination of the light element abundances has an impact on our exclusions.
Recall that the coupling $\tilde{A}\rightarrow \tilde{S}\tilde{S}$ is proportional to $\lambda_{S12}-\lambda_{S21}$, therefore points excluded by  BBN constraints  have $\lambda_{S12}-\lambda_{S21}< 10^{-13}$, see Fig.~\ref{fig:param_A}(left).  Moreover  since the decay  $\tilde{A}\rightarrow  Z \tilde{S}\tilde{S}$  is  $\propto \lambda_{S12}+\lambda_{S21}$, it becomes dominant over the two-body decay when $\lambda_{S12}\approx \lambda_{S21}$.  
Note that to compute the yield of the doublet, we solve  Eq.~\ref{eq:Y2} at $T =10^{-3} {\rm GeV}$.   The impact of the BBN constraint on the parameter space of the model is shown in Fig~\ref{fig:param_A}. In particular since as mentionned above that BBN constrain small values of $\lambda_{S12}$ and $\lambda_{S21}$, it implies that BBN also constrains "large" values of $\lambda_{S1}$ or $\lambda_{S2}$, indeed one of these couplings must be in the range $10^{-12}-10^{-9}$  to produce enough DM formation through freeze-in.
 The main constraint on the doublet mass comes from  the upper limit on the Higgs invisible width which rules out the region where $M_A< 62 {\rm GeV}$. On the other hand 
 the decay of the Higgs to DM pairs is extremely small and cannot be constrained from measurements of the Higgs invisible width, we find that because of the feeble couplings involved, $Br(h\to \tilde{S}\tilde{S})< 10^{-15}$.  A possible signature of  this scenario at the LHC is the monojet signal. However we  show in Fig.~\ref{fig:LHC_A} that the current and projected monojet limits, only  have a marginal impact at low doublet masses. The theoretical constraints on the value of $\lambda_{Ah}$ are such that the monojet signal is suppressed at masses above the EW scale as in the IDM. Other collider signatures of this scenario are  similar to those of the inert doublet model when $\tilde{A}$ decays invisibly into pairs of singlet DM. Results obtained in the IDM for the same-sign dilepton searches in Ref.~\cite{Yang:2021hcu}  or for the reach of  future colliders ~\cite{Kalinowski:2020rmb} can be applied here. 
A  distinctive signature of this scenario could be associated with the process $pp\rightarrow \tilde{A}\tilde{H}\rightarrow (Z\tilde{S}\tilde{S})(\tilde{A} Z)$ leading to two Z's and missing energy. However we found that  when the decay $\tilde{A}\rightarrow Z^{(*)} \tilde{S}\tilde{S}$ has a large branching fraction  the lifetime of $\tilde{A}$ is long enough that the decay occurs outside the detector.  
The same process, when $\tilde{A}$ is long-lived could be detectable with MATHUSLA. However we found that the cross-section times branching ratio were several orders of magnitude below the reach, see Fig.~\ref{fig:LHC_A} for $\tilde{A}\tilde{H}$ production. Here we have used the approximate formula in Ref.~\cite{Curtin:2018mvb} to estimate the boost factor $b$. Note that the cross-section for $\tilde{A}\tilde{A}$ is even more suppressed. 
  Under these conditions the singlet plays little role in collider searches, we however stress that since the singlet allows to fulfill the relic density constraint, the regions where the IDM does not satisfy DM constraint should be explored at colliders due to the presence of the FIMP.

\newpage
\subsection{The case where $M_{H+} < M_A,M_H$}
For this scenario, we performed a dedicated scan choosing for  the  range of values for the feeble couplings  $10^{-15} - 10^{-9}$  while  the masses are varied in the range
\begin{equation}
M_S= 1-500 {\rm GeV}, \;\;\;  2M_S <  M_{H^+}  <1000 {\rm GeV},\;\;  M_{A/H}-M_{H^+} < 500 {\rm GeV}.
\end{equation} 
We do not consider  $M_{H+}<500GeV$, since we  verified that  all these points  are  ruled out by HSCP searches at the LHC. Indeed, the charged Higgs always decay outside the detector  since its dominant  decay mode  $~\tilde{H}^+\rightarrow W^+\tilde{S}\tilde{S}$  is suppressed by the small couplings.

\begin{figure}[h]
		\includegraphics[scale=0.5]{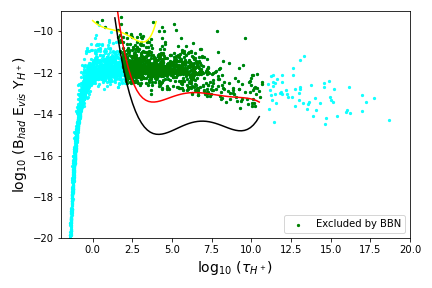}
		\includegraphics[scale=0.5]{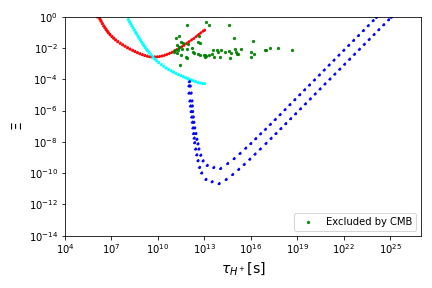}			
			\caption{Points allowed (cyan) and ruled out (green) by BBN in the $B_{had} \epsilon_{had} Y_A$ vs $\tau(H^+)$ plane (left), exclusion limits as in Fig.~\ref{fig:BBN} (left). The impact of CMB constraints for scenarios where  $\tau_{H+}>10^{12}$ sec in the plane $\Xi$ - lifetime(right), the exclusion limits are from Ref. ~\cite{Poulin:2016anj},  the red and cyan curve  correspond to spectral distorsion constraints from FIRAS while the blue dots are the limits from anisotropies of the power spectra (including uncertainties). 			 }   
	\label{fig:BBNcharged} 
\end{figure}

\begin{figure}[h]
\begin{center}
\includegraphics[scale=0.5]{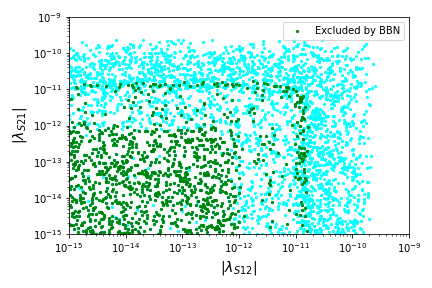}	
\end{center}	
\caption{Points allowed (cyan) and ruled out (green) by BBN and CMB  in the  $\lambda_{S12}- \lambda_{S21}$ plane. }   
\label{fig:couplings_CH} 
\end{figure}

As discussed in the previous section, the singlet can freeze-in through its interactions with the SM (which requires non-negligible $\lambda_{S1}$) or through its interactions with the doublet which requires either $\lambda_{S2}, \lambda_{S12}$ or $\lambda_{S21}$ to be $\approx 10^{-11}-10^{-10}$. Moreover the decay of the lightest doublet component can occur before or after the freeze-out of the charged Higgs through the process $\tilde{H}^+\rightarrow W^+ \tilde{S}\tilde{S}$. In general all charged Higgs have  decayed today and do not contribute to the DM density,  however in a few cases where $M_{H+}\approx 2M_S$ the decay of the charged Higgs is strongly  suppressed and its abundance is not reduced, thus $\Omega_2 h^2\approx 0.1$. These scenarios are eliminated since a stable charged particle is subject to various constraints and is ruled out for the range of masses we consider even if it forms a small fraction of the total DM, see Ref.~\cite{Dunsky:2018mqs}.    In addition we find  that all lifetimes of the charged Higgs greater than 100sec are ruled out by either BBN or CMB constraints, see Fig.~\ref{fig:BBNcharged}. CMB constraints  from Ref.~\cite{Poulin:2016anj} are applied as described in Sec.~\ref{sec:CMB} for scenarios where $\tau> 10^{12}$ sec.  We have estimated the electromagnetic energy injection from the decay $\tilde{H}^+\rightarrow W^+ \tilde{S}\tilde{S}$ to be roughly $E_{em}=40\%$, a more precise estimate is clearly not necessary as $\Xi$ exceeds the allowed  values by several orders of magnitude. 
Since the lifetime depends on $\lambda_{S12},\lambda_{S21}$ we find that these couplings need generally to be  below $10^{-12}$ for the scenario to be ruled out by BBN, see Fig.~\ref{fig:couplings_CH}. In cases where $M_{H+}-2M_S<M_W$ such that the main decay process is 4-body, couplings of the order $10^{-11}$  can also be excluded by BBN.
 
At colliders, this scenario is peculiar and distinguishable from the inert doublet model  since the lightest doublet component is charged. Improved searches for HSCP therefore provide the most promising searches for this scenario since the lifetime of $\tilde{H}^+$ is always very long with $c\tau>10^5 m$. Apart from the pair production of $\tilde{H}^+$, other signatures involving a stable charged particles include $pp\rightarrow \tilde{H}^+ \tilde{A} (\tilde{H})$ which leads to a stable charged particle and missing transverse energy signature when  $\tilde{A}$ or $\tilde{H}$ are either long-lived at the collider scale or decay into SS. Alternate signatures correspond to  a stable charged particle and a Z when $\tilde{A}(\tilde{H}) \rightarrow \tilde{H}(\tilde{A})Z$ or two stable stable particles and a W when  $\tilde{H}\rightarrow W^\pm \tilde{H}^\mp$. Finally as in the IDM, $HA$ production could lead to a monoZ signature. However, as in the IDM,  production cross-sections are at best ${\cal{O}}(0.1)$ fb level for the heavy spectra we consider and therefore these signatures are  more adapted to higher energy versions of the LHC~\cite{Kalinowski:2020rmb,Robens:2021zvr}.

\section{Conclusion}

We have shown that in the inert doublet and singlet model, a singlet DM that is feebly interacting can be the dominant DM component. In this case, astroparticle or collider searches for DM are sensitive only to the doublet sector. Because the doublet could only be a small fraction of DM, it could escape both direct and indirect searches. Thus collider searches  could provide the only evidence for new physics and are complementary to astroparticle searches. Many of  the scenarios that are currently allowed by all constraints can lead to signatures of long-lived particles or of disappearing tracks. It remains to be seen how well the scenarios can be probe with the future HL-LHC. These signatures are especially important in the sense that most of the scenarios that escape future direct detection searches predict a lifetime of the charged Higgs around $c\tau\approx 10^{-4} - 1 {\rm m}$ covering the region with displaced signatures. 
As typically found for FIMPs, the singlet can be well below the electroweak scale.

We have also investigated scenarios where the singlet FIMP is the only DM component. After taking into account BBN constraints, when the neutral Higgs is the lightest doublet component we find that the model shares many characteristics of the inert doublet model apart from the fact that the whole range of masses is allowed. Collider searches should therefore not be restricted to the ranges around 60 GeV or above 500GeV which correspond to DM compatible with the relic density in the IDM. When the charged Higgs is the lightest doublet component, its lifetime can be long enough that cosmological constraints apply as well. The distinctive signature of the model at colliders corresponds to HSCP and current searches require the charged scalar mass to be above 550 GeV, upgrades of the LHC will be able to cover higher masses of the charged Higgs.

\section{Acknowledgements}

We thank Christopher Eckner for helpful discussions on CTA and the use of his code for reading the spectrum tables. We also acknowledge useful discussions with Alexander Belyaev, Fawzi Boudjema and Pasquale Serpico.
This work  was funded by RFBR and CNRS, project number 20-52-15005. The work of A. Pukhov was also supported in part by a grant AAP-USMB in 2021.

\providecommand{\href}[2]{#2}\begingroup\raggedright\endgroup

\end{document}